\documentclass[aps,prl,twocolumn,superscriptaddress,showpacs,floatfix]{revtex4-1}
\usepackage{amsmath,amssymb,multirow,epsfig,bm,dcolumn,pstricks}

\usepackage{ulem}

\begin{document}
\title{Cold neutrons trapped in external fields
}

\author{S. Gandolfi}
\affiliation{ Theoretical Division,
Los Alamos National Laboratory,
Los Alamos, NM 87545}

\author{J. Carlson}
\affiliation{ Theoretical Division,
Los Alamos National Laboratory,
Los Alamos, NM 87545}

\author{Steven C. Pieper}
\affiliation{ Physics Division,
Argonne National Laboratory,
Argonne, IL 61801}

\begin{abstract}
The properties of inhomogeneous neutron matter are crucial to the
physics of neutron-rich nuclei and the crust of neutron stars.  Advances in
computational techniques now allow us to accurately determine the binding
energies and densities of many neutrons interacting via realistic microscopic 
interactions and confined in external fields.  We 
perform calculations for different external fields and across several
shells to place important constraints on inhomogeneous neutron
matter, and hence the large isospin limit of the nuclear energy density functionals
that are used to predict properties of heavy nuclei and neutron star crusts.
We find important differences between microscopic calculations and current
density functionals; in particular the
isovector gradient terms are significantly more repulsive than in
traditional models, and the spin-orbit and pairing forces are
comparatively weaker.  
\end{abstract}

%\date{\today} 

% 21.10.-k      Properties of nuclei; nuclear energy levels
% 21.30.-x 	Nuclear forces
% 21.60.-n 	Nuclear structure models and methods
% 21.60.De 	Ab initio methods 
% 21.60.Jz 	Nuclear Density Functional Theory and extensions

\maketitle 

The properties of inhomogeneous neutron-rich
matter are important in both astrophysical and terrestrial regimes.
While the equation of state (EOS) and the pairing gap for homogeneous neutron
matter have been studied extensively in microscopic theories
\cite{APR:1998,Dean:2003,Gandolfi:2009,Gezerlis:2008:2010},
inhomogeneous neutron matter has received comparably little
attention. Understanding the inner crust of neutron stars, which
affects transient stellar cooling and determine
oscillation modes requires  knowledge
of inhomogeneous neutron-rich
matter~\cite{Ravenhall:1983,Shternin:2007,Brown:2009}.
Neutron-rich nuclei are
also the subject of intense theoretical and experimental investigations,
driven by their relevance for r-process nucleosynthesis as well as the
intrinsic interest in the properties of nuclei at large isospin;~\cite{Dobaczewski:1996,Dobaczewski:2007}
they are the principal thrust of rare-isotope accelerators~\cite{Geesaman:2006}.

Simulations of both the crust of neutron stars and of large
neutron-rich nuclei employ nuclear energy density functionals fit to 
nuclei.  These density functionals have proved to be
extremely successful in describing many nuclei, but involve large extrapolations
to reach inhomogeneous neutron matter.  To test these
extrapolations, we perform calculations of neutron drops --
neutrons confined by artificial external fields and interacting via
realistic two- and three-nucleon forces.  
We vary
substantially the number of neutrons as well as the strength and shape of
the external fields to test the density functional.
  
The EOS of homogeneous neutron matter has often been
included as a constraint to density functional theories (eg. \cite{Chabanat:1998}); our objective
is to allow inhomogeneous neutron matter to be employed in a similar
manner.  We find, for example, that once the bulk terms are fixed from the neutron matter
EOS, the closed shells of neutrons are
primarily sensitive to the gradient terms in the density functional.
These
pure neutron matter gradient terms have modest effects on nuclei, and hence they are not well
constrained in fits to nuclear masses \cite{Furnstahl:2000,Brown:1998}.  
The closed-shell systems are
nearly independent of spin-orbit and pairing terms, but ground and excited
states of a single neutron outside a closed shell, 
or of a single neutron hole, are a sensitive probe
of the spin-orbit interaction.  Mid-shell results are
sensitive to both spin-orbit and pairing terms.
We compare our calculated results to several ``standard'' Skyrme models,
and also to a model in which the isovector terms are
adjusted to reproduce the ab-initio calculations; these changes are expected
to have only a small effect on the nuclear energies used to fit the original 
parameters.  The goal of
these studies is to determine which terms in the
density functional can be probed through microscopic
calculations, and how the adjusted values compare to traditional models.
A realistic improved density functional will require a complete
refitting of nuclear properties along with the properties of homogeneous
and inhomogeneous neutron matter~\cite{unedf,Bertsch:2005}.

{\it Interaction and Methods:} We report calculations of
neutrons in harmonic oscillators (HO) of two frequencies and a Woods-Saxon (WS)
well.  The full Hamiltonian is:
\begin{equation}
H =  -\frac{\hbar^2}{2 m} \sum_i  \nabla_i^2 + \sum_i V_i + \sum_{i<j} V_{ij} + \sum_{i<j<k} V_{ijk}, \nonumber
\end{equation}
where $V_i = ( m \omega^2 / 2 ) r_i^2$ (HO) or $V_i = -V_0 / ( 1 + \exp [ (r_i -
r_0)/a])$ (WS) with $V_0 = -35.5$ MeV, $r_0=3$ fm and $a=1.1$ fm, and $ \hbar^2 / m = 41.44$ MeV-fm$^2$.  The
neutron-neutron potential $V_{ij}$ is 
AV8$^\prime$~\cite{Pudliner:1997}, a slightly simplified version of the AV18 potential~\cite{Wiringa:1995};
we find less than 0.25\% differences in neutron-drop energies for these two potentials.  We also add the
Urbana IX model (UIX)~\cite{Pudliner:1997} three-nucleon interaction (TNI), including
the p-wave two-pion exchange (Fujita-Miyazawa) TNI and a short-range phenomenological
repulsion.
%or JISP16\cite{Shirokov:2007}. 
We use 
this combination of two- and three-nucleon interactions because it produces an
EOS consistent with known neutron star masses~\cite{APR:1998},
and because several present-day Skyrme models have used this
EOS to constrain the properties of homogeneous neutron matter.
Further studies with different interaction models will be valuable, in particular to look
at the spin-orbit interactions which might be increased with a three-pion
exchange TNI as in Illinois-7~\cite{Pieper:2008}.

\begin{figure}
\includegraphics[width=3.2in]{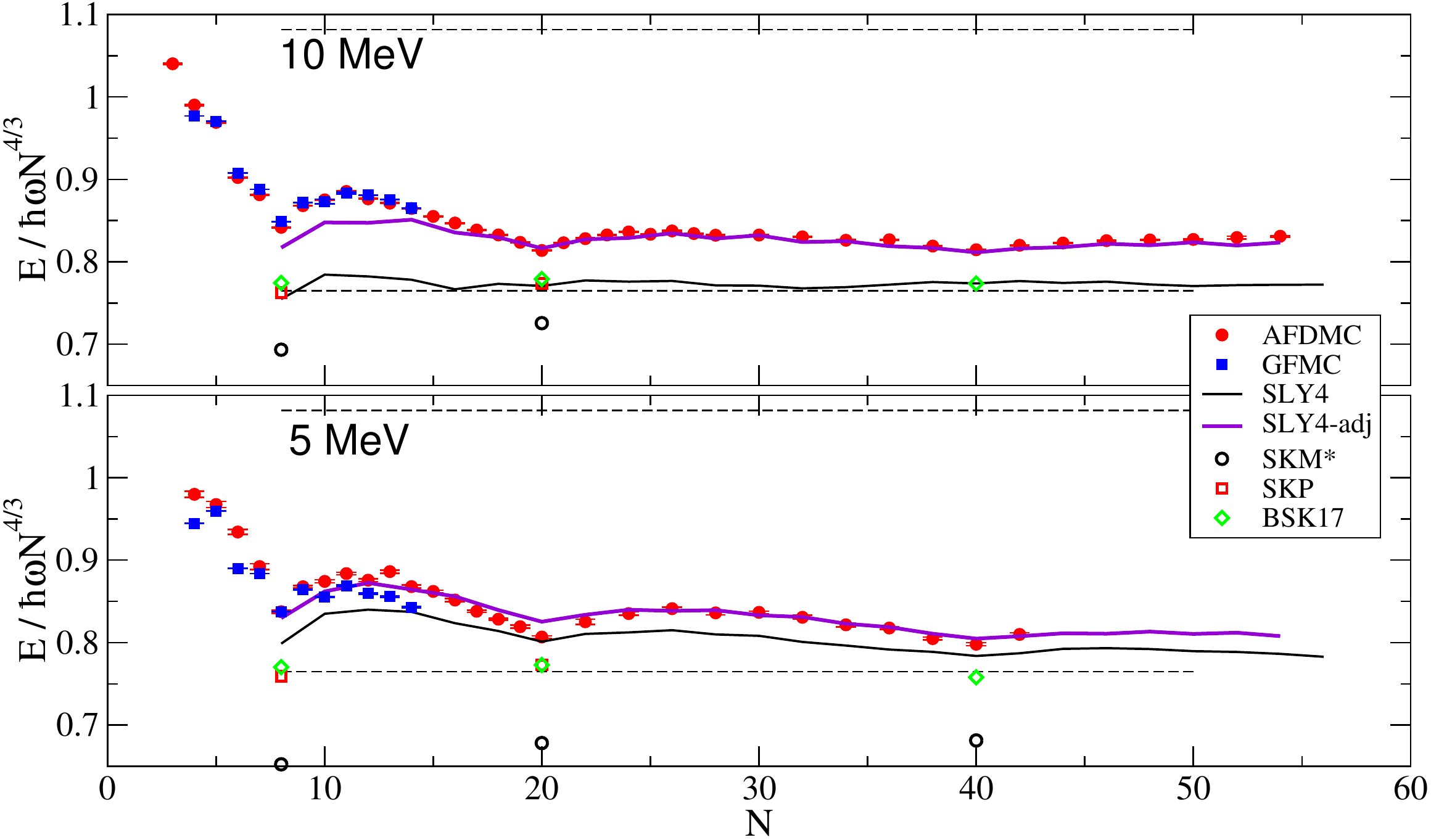}
\caption{(color online) Energies divided by $\hbar\omega N^{4/3}$ for neutrons in 
HO fields with $\hbar\omega = 10$~MeV (top) and 5~MeV
(bottom).  Filled symbols indicate ab initio calculations;
the dashed lines are Thomas-Fermi results (see text);
the lower curves
are from the SLY4 interaction and the upper curves show
the modified SLY4 interaction described in the text.
}
\label{fig:hospectra}
\end{figure}

Calculations are performed using Green's Function Monte Carlo 
(GFMC)~\cite{PWC04} and Auxiliary Field 
Diffusion Monte Carlo (AFDMC)~\cite{Schmidt:1999} quantum Monte Carlo (QMC)
methods.
These algorithms evolve an initial trial state, $\Psi_T$, in
imaginary time to yield the ground-state.  GFMC sums explicitly over
spin and isospin states, and can use very sophisticated  
$\Psi_T$~\cite{Pudliner:1997}.  
However it
is limited to small systems, up to 16 neutrons.
%  AFDMC samples spin and isospin degrees of freedom as well as the
%  spatial integrals, and hence it can treat larger
In addition to sampling the spatial integrals as in GFMC, AFDMC also
samples the spin and isospin degrees of freedom, and hence it can treat larger
systems~\cite{Gandolfi:2009}.  Both methods use a constraint
involving the overlap with $\Psi_T$ to eliminate the Fermion sign
problem, and hence are approximate.  
Studies of light nuclei and neutron matter show they give results
within $1\%$ of the exact ground-state energy.

\begin{figure}
\includegraphics[width=3.2in]{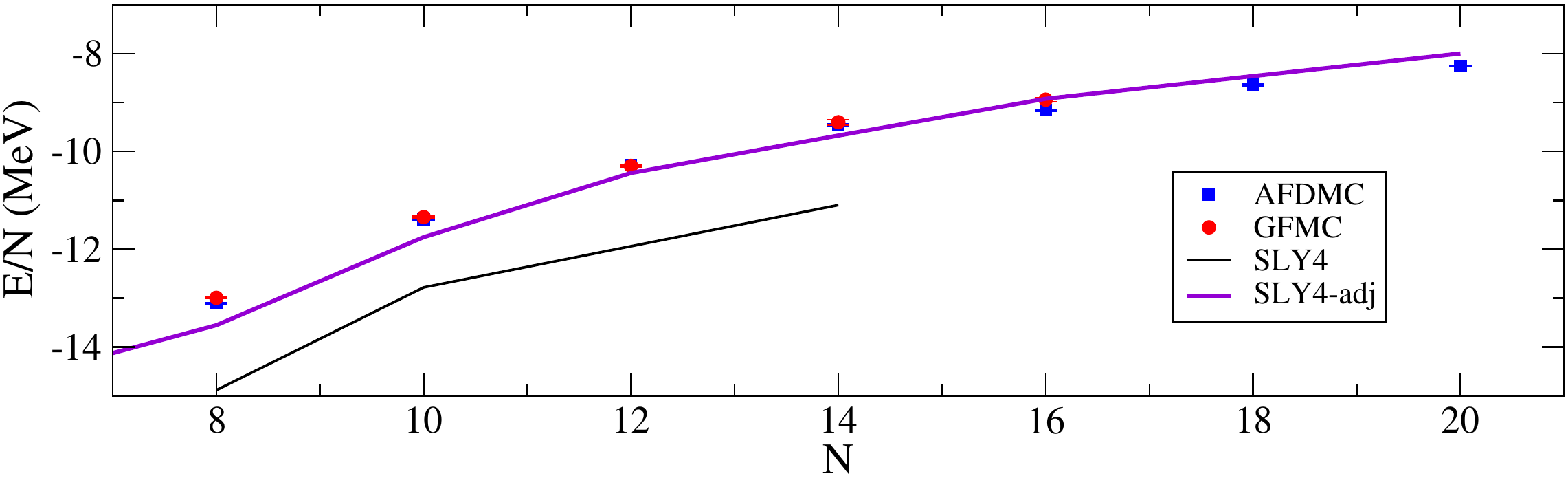}
\caption{(color online) Energies per particle for neutrons in the Woods-Saxon field, symbols as in Fig.~\ref{fig:hospectra}.
}
\label{fig:wsspectra}
\end{figure}

We use external fields yielding
low or moderate densities. 
However, even at small densities neutrons are strongly
interacting  and pairing can be important.
Recent microscopic
calculations of neutron matter give s-wave
pairing gaps of several MeV~\cite{Gezerlis:2008:2010, Gandolfi:2009a}.
One- and two-nucleon properties including pairing gaps and spin-orbit splittings
can be more sensitive to models of the three-nucleon interaction.
Calculations of very small neutron drops (N=6,7,8) have been performed
previously~\cite{Pudliner:1996,Smerzi:1997,Pederiva:2004}. Even these
calculations indicated a substantial difference with traditional Skyrme
models, which overbind the drops and give too-large spin-orbit splitting.

{\it Results:} The ground-state energies versus neutron number N 
for the HO potentials are given in
Fig.~\ref{fig:hospectra} and for the WS potential in Fig.~\ref{fig:wsspectra}.  
Up to N=16 both
GFMC and AFDMC results are included. They agree very well for the 10-MeV
HO interaction, while for $\hbar\omega = 5$~MeV, the AFDMC results are
slightly higher than the GFMC; the maximum difference is $3\%$, and more
typically results are within $1\%$. The bigger difference for the lower
density 5-MeV drops presumably arises because the
AFDMC $\Psi_T$ does not yet include pairing, while the GFMC does.

\begin{table}
\begin{tabular}[c]{rcdddd}
 \hline
 \hline
N & $J^{\pi}$ & \multicolumn{2}{c}{$\hbar \omega$ = 5 MeV} & 
\multicolumn{2}{c}{$\hbar \omega$ = 10 MeV} \\
 & & \multicolumn{1}{r}{\rm GFMC}& \multicolumn{1}{r}{\rm AFDMC} & 
     \multicolumn{1}{r}{\rm GFMC} & \multicolumn{1}{r}{\rm AFDMC} \\
 \hline
 7 & $1/2^-$ & 59.17(1) & 59.7(2) & 118.95(3) & 118.01(9) \\
 7 & $3/2^-$ & 59.73(1) & 60.3(2) & 121.08(3) & 120.57(7)\\
% \vspace*{-6pt}\\
 8 & $0^+$  &  67.01(1) & 67.0(2) & 135.76(4) & 134.7(1)  \\
% \vspace*{-6pt}\\
 9 & $5/2^+$ & 81.20(3) & 81.6(2) & 163.2(1)  & 162.5(1) \\
 9 & $3/2^+$ &          & 82.3(2) &           & 166.1(1) \\
% \vspace*{-6pt}\\
 10 & $0^+$ &  92.1(1) & 94.2(2) &  188.1(6) & 188.5(1) \\
% \vspace*{-6pt}\\
 12 & $0^+$  & 118.1(1)  & 120.3(3) &  242.0(6)  & 240.8(1)  \\
% \vspace*{-6pt}\\
 13 & $5/2^+$ & 131.5(1) & 135.4(3) &  267.6(6)  & 266.3(2) \\
 13 & $3/2^+$ &          &          &            & 269.3(2)  \\
% \vspace*{-6pt}\\
 14 & $0^+$  &  142.2(1) & 146.4(3) &  291.9(2)  & 291.7(2) \\
% \vspace*{-6pt}\\
 20 & $0^+$  &           & 219.0(4) &            & 441.7(4) \\
% \vspace*{-6pt}\\
 40 & $0^+$  &           & 545.8(1.3)&            &1114.3(9) \\
 \hline
 \hline
 \end{tabular}
\caption{ Selected energies from GFMC and AFDMC calculations using
AV8$^\prime$+UIX with HO external fields. }
 \label{tab:spinorbit}
 \end{table} 

In addition to the microscopic calculations, results for several different
Skyrme models are shown in Fig.~\ref{fig:hospectra}. We also show results
for Thomas-Fermi local density approximations~\cite{Chang:2007} using $E(\rho_n)/N
= \xi (3/5) (\hbar^2/2m) ( 3 \pi^2 \rho_n)^{2/3}$; the upper horizontal line is for
free particles, $\xi = 1$, and the lower has $\xi = 0.5$, a reasonable
approximation to the EOS of low-density neutron matter.
For the 10-MeV well, the density functionals give energies
significantly below the Monte Carlo results for all N.
The energies are also lower for the 5-MeV well, but less so. 
This overbinding is a general feature of all the Skyrme models considered.
It is intriguing that these same Skyrme models underbind the 
properties of very dilute neutron systems, typically they are fit
to the neutron matter EOS at $\rho = 0.04 $ fm$^{-3}$ and above.

Since the Skyrme homogeneous neutron matter EOS
have been fit to various microscopic calculations,
this overbinding suggests that the gradient
terms in  inhomogeneous neutron matter should be more repulsive.
The observed differences between ab-initio results and the
Skyrme functionals are much larger than the differences 
between experiments and Skyrme models in nuclei, as expected,
because of the large extrapolations to inhomogeneous neutron matter.

{\it Isovector Gradient Contributions:} As is apparent in Fig.~\ref{fig:hospectra},
for harmonic oscillators there are closed shells at N= $8, 20,$ and $40$
neutrons.  These closed-shell states are almost exclusively sensitive to
the neutron matter EOS and the isovector gradient terms;
pairing and spin-orbit play nearly no role. Hence they are
direct probes of the gradient terms; to examine them we have altered the
isovector gradient terms in the SLY4 interaction~\cite{Chabanat:1998} to
approximately reproduce the QMC results using a modified version of the
ev8 code~\cite{Bonche:1985}, The gradient terms are adjusted
without changing any isoscalar (T=0) parameters or the homogeneous neutron
matter EOS.

The lowest-order gradient contribution to the energy density
for inhomogeneous matter is $G_d [ \nabla \rho_n ]^2$.
The constants $G_d$ are small
and often negative, for example, $G_d = -16, -7, 17, -17, -7$
MeV-fm$^{5}$ for the SLY4, SLY7, BSK17, SkM$^*$, and SkP
interactions.  Repulsive gradient terms for neutron matter are to be
expected on rather general grounds, and are required for the absolute
stability of uniform matter in the absence of a background field.  The
adjusted interaction SLY4-adj gives $G_d = 26.5$, a similar
adjustment to the BSK17 interaction which is more attractive for
homogeneous neutron matter yields $G_d = 64$.  A single adjustment
of $G_d$ markedly improves the agreement with QMC results for both the
HO and WS fields. A
precise fit to both neutron matter and these results would require a
more general form of the density functional.  

{\it Isovector Spin-Orbit:} By examining neutron numbers slightly away
closed shells, we can constrain the spin-orbit interaction for neutron
drops.  For example, $N=7,9$ results are sensitive to the spin-orbit
interaction, but not to the pairing terms.  We find the spin-orbit
interaction to be small in the calculated drops; the energies 
for some low-lying spin-orbit partners are given in Table
\ref{tab:spinorbit}.

Small spin-orbit splitting had been found previously in calculations of N=7 drops and our results
show similar effects for all systems near closed shells
(N=7,9,11,...).  The simplest (standard) Skyrme parametrizations
give a strength ratio of 3:1 between isoscalar ($T=0$) and isovector
($T=1$) spin-orbit couplings.  We find an even smaller isovector coupling,
approximately 1/6 of the isoscalar coupling,
reproduces the ab initio calculations.
The combined factor of $1/6$ is in reasonable
agreement with $1/8$ found in a diagrammatic examination of
spin-orbit splittings from microscopic interactions~\cite{Kaiser:2004},
and with results obtained in an examination of Skyrme parameters from Pb
isotopes~\cite{Reinhard:1995}. Relativistic mean fields yield zero
strength in the isovector channel~\cite{Onsi:1997}, while recent results
from the pion contribution from chiral interactions give nearly 
equal isovector and isoscalar strengths~\cite{Kaiser:2010}.

{\it Isovector Pairing:} The mid-shell results (eg. N=14, 30) and
odd-even staggerings are sensitive to the pairing interactions as well
as the spin-orbit force.  Fixing the spin-orbit strength from near
closed-shells, we adjust the pairing strength to fit the
calculated spectra.  There is a significant interplay between
the pairing and spin-orbit forces required to reproduce microscopic calculations.  A small spin-orbit force results in many quasi-degenerate
levels which enhances pairing in mid-shell systems.

Several models of pairing are used in
density-functional theories.  We employ a simple volume parametrization
with a delta-function spatial dependence, a density cutoff that
restricts pairing to $\rho_n < \rho_0$, 
and limit the pairing to single-particle orbitals less than 
5 MeV from the Fermi energy. 
We find a reduction from a typical 1 GeV-fm$^3$ strength
to half that value significantly improves agreement with microscopic
results.  A reduction of pairing in neutron-rich
nuclei has recently been found to give a better fit to experimental
energy differences of 156 nuclei of mass A=118 to
196~\cite{Yamagami:2009}.

Adjusting these three parameters 
(gradient term $G_d$ = 26.5, spin-orbit coupling = 123 MeV-fm$^3$ and pairing strength = 500 MeV)
in the density functional 
increases the
agreement across all external fields and all particle numbers. 
This is shown by the upper solid curves (SLY4-adj) in Fig.~\ref{fig:hospectra}.

\begin{figure}
\includegraphics[width=3.2in]{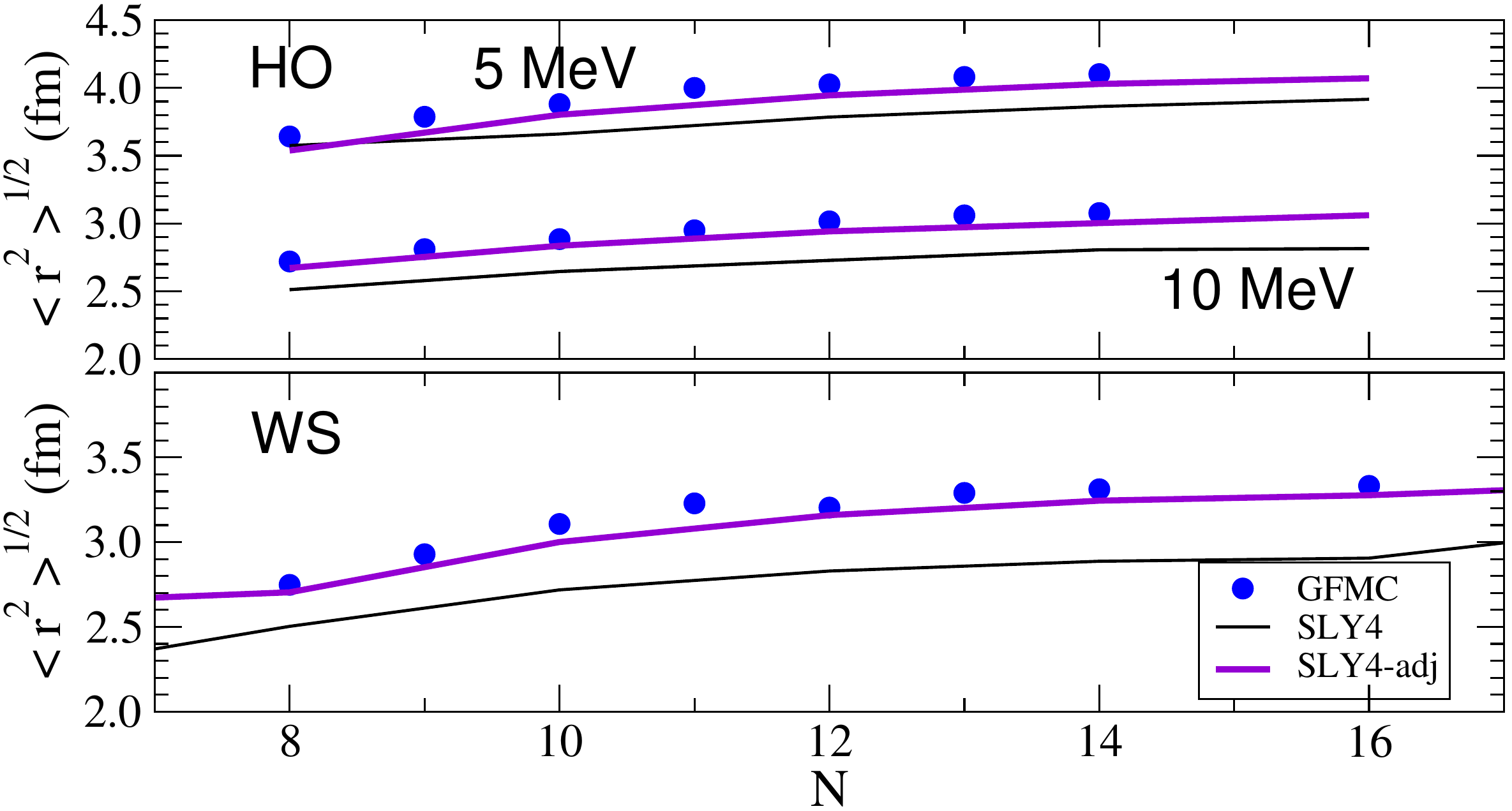}
\caption{(color online) Calculated radii of neutrons confined in HO
(upper) and WS (lower), fields compared to original and adjusted
Skyrme models (see text).}
\label{fig:horadii}
\end{figure}

{\it Radii and Mass Distributions:} Our calculations yield precise
estimates for RMS radii and the density distributions of the smaller drops. 
The average densities, defined as
$\int d^3r \rho_n^2(r) / N$,
of the drops in the 5-MeV HO well are
approximately 0.02 fm$^{-3}$, or
about 1/8 nuclear matter saturation density, while for the 10-MeV HO
and the WS wells
they are $\sim$ 0.045 fm$^{-3}$, or almost 1/3 nuclear matter saturation density.

\begin{figure}
\label{fig:dist}
\includegraphics[width=3.2in]{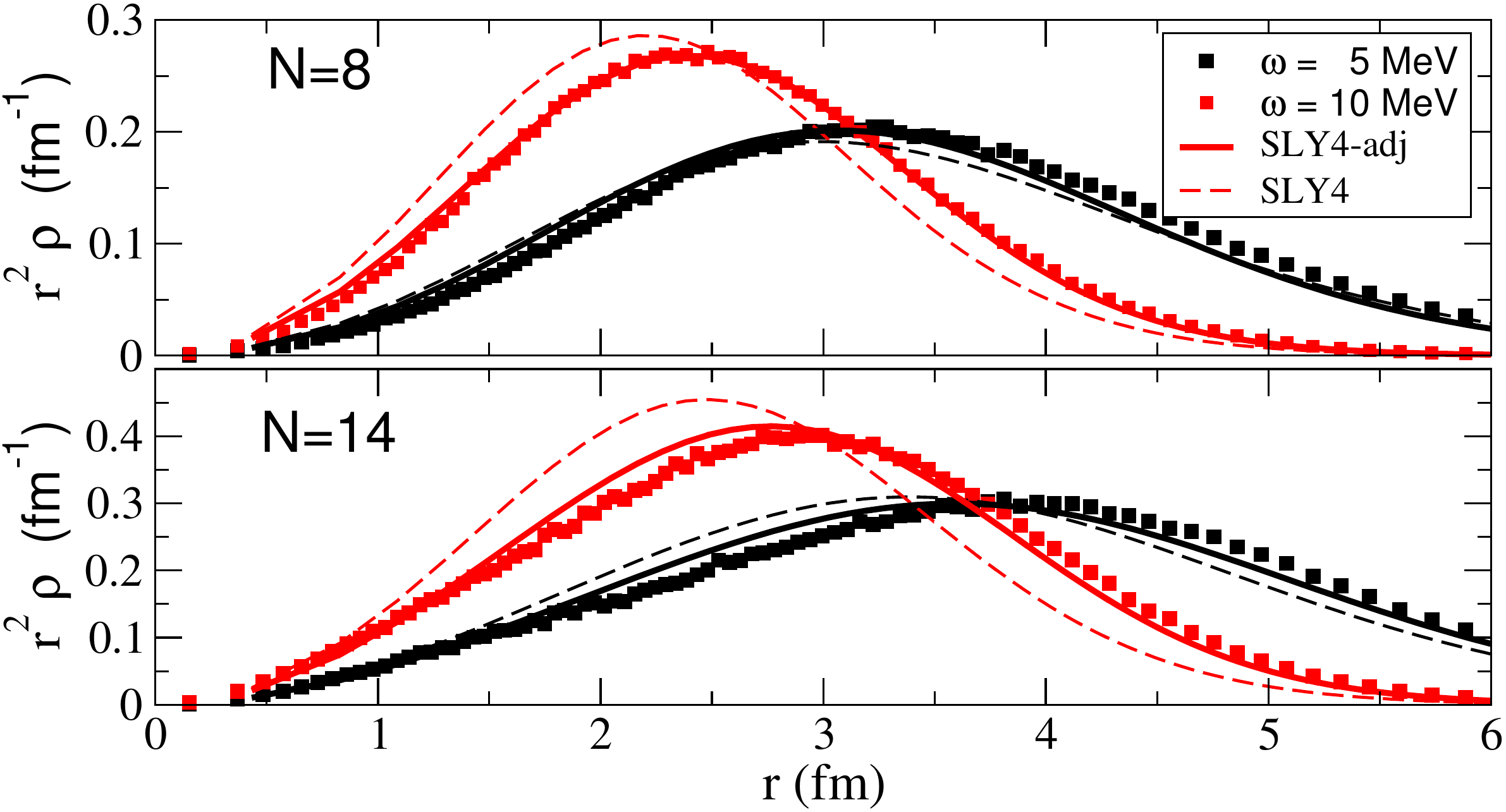}
\caption{(color online) Calculated densities of neutrons in HO
potentials, compared to Skyrme models (see text).}
\label{fig:hodist}
\end{figure}

The RMS radii obtained in microscopic calculations are compared with the
original and adjusted Skyrme density functional results in Fig.~\ref{fig:horadii}.
The density distributions for N=8 and 14 are compared in Fig.~\ref{fig:hodist}.
Since we are comparing
gross properties of inhomogeneous matter,
we plot the densities weighted with the phase
space: $r^2 \rho_n (r)$, which gives a better picture of the density distributions
near the average density of the system.
In every case the adjusted Skyrme interaction produces a better description
of the radii and density distributions.  The $N=8$ calculations depend
primarily upon the gradient terms, the reduction in pairing and spin-orbit
are also important for $N=14$. The improvement in the mid-shell N=14
case is particularly dramatic, as a significant shift in the density occurs
with the modified isovector Skyrme parameters, bringing the results
into much better agreement with microscopic calculations.

{\it Conclusions:}
We have examined the properties of neutrons confined in external fields
to study the properties of inhomogeneous neutron matter.  These ab-initio calculations
place significant constraints on the nuclear energy density functional
in a regime far from that probed by fitting to available nuclei.  They indicate the need
for more repulsive gradient terms in pure neutron matter, and a reduced 
isovector spin-orbit and pairing strength compared to standard functionals.
With a combined fit of density functionals to both nuclei
and neutron matter, more reliable predictions should be possible for very
neutron-rich nuclei including those participating in r-process
nucleosynthesis.  These improved functionals would also be extremely
valuable in examining the neutron skin thickness of
lead~\cite{Reinhard:2010}, as can be probed in parity-violating electron
scattering.  Much more reliable predictions for extremely neutron-rich
astrophysical environments can also be expected.
The numerical values of the results shown in the figures are given
in Ref.~\cite{epaps}.

We thank G. F. Bertsch, A. Bulgac, S. a Beccara,
J. Dobaczewski, W. Nazarewicz, P. Maris, F. Pederiva, S. Reddy, J. Vary, 
and R. B. Wiringa  for valuable discussions.  We are indebted to K. E. Schmidt for providing us the
AFDMC code. 
This work is supported by the U.S. Department of Energy,
Office of Nuclear Physics, under contracts DE-FC02-07ER41457 
(UNEDF SciDAC), DE-AC02-06CH11357,
and DE-AC52-06NA25396. Computer time was made
available by Argonne's LCRC, the Argonne Mathematics and
Computer Science Division, Los Alamos Open Supercomputing, 
the National Energy Research Scientic Computing
Center (NERSC) and by a DOE INCITE grant on the
Argonne BG/P.

\bibliographystyle{apsrev4-1}

\end{document}